\documentclass[prl,final,twocolumn,superscriptaddress]{revtex4-1}
\usepackage{amsmath}
\usepackage{graphicx}

\usepackage[usenames,dvipsnames]{color}
\usepackage{enumitem}

\newcommand{\mx}{Mn$_3$X }
\begin{document}

\title{ 
 Dzyaloshinskii-Moriya interaction in absence of spin-orbit coupling
}




\author{Ramon Cardias}
\affiliation{Faculdade de F\'\i sica, Universidade Federal do Par\'a, Bel\'em, PA, Brazil}
\affiliation{Department of Physics and Astronomy, Uppsala University, 75120 Box 516 Sweden}
\affiliation{SPEC, CEA, CNRS, Université Paris-Saclay, CEA Saclay, 91191 Gif-Sur-Yvette, France}

\author{Anders Bergman}
\author{Attila Szilva}
\author{Yaroslav O. Kvashnin}
\author{Jonas Fransson}
\affiliation{Department of Physics and Astronomy, Uppsala University, 75120 Box 516 Sweden}

\author{Angela B. Klautau}
\affiliation{Faculdade de F\'\i sica, Universidade Federal do Par\'a, Bel\'em, PA, Brazil}

\author{Olle Eriksson}
\affiliation{Department of Physics and Astronomy, Uppsala University, 75120 Box 516 Sweden}
\affiliation{School of Science and Technology, \"Orebro University, SE-701 82, \"Orebro, Sweden}

\author{Lars Nordstr\"om}
\affiliation{Department of Physics and Astronomy, Uppsala University, 75120 Box 516 Sweden}

\date{\today}

\begin{abstract}
In contrast to conventional assumptions, we show that the Dzyaloshinskii-Moriya interaction can be of non-relativistic origin. In materials with a non-collinear magnetic configuration, the non-relativistic contributions can dominate over the contribution due to spin-orbit coupling. The weak antiferromagnetic phase of the co-planar state of  Mn$_{3}$Sn is found to be due to a combinaton of this non-relativistic and the relativistic contribution to the Dzyaloshinskii-Moriya interaction. Using electronic structure theory as a conceptual platform, all relevant exchange interactions are derived for a general, non-collinear magnetic state. It is demonstrated that non-collinearity influences all three types of exchange interaction and that physically distinct mechanisms, which connect to  electron- and spin-density and currents, may be used as a general way to analyze and understand magnetic interactions of the solid state. 

\end{abstract}

\maketitle

\section{Introduction}
The search for spintronic devices, where the electron spin is in focus - in contrast to the field of electronics that rely on the electron charge, has recently turned focus on antiferromagnets \cite{Jungwirth:2018if,Smejkal:2018ew}. 
There are many advantages with antiferromagnetic spintronic. Besides the nearly disipationless transport without heat production, the spin dynamics of an antiferromagnet is significantly faster than that of a ferromagnet, and simultaneously there are less problems with magnetic stray fields.
Among the class of promising antiferromagnets, triangular Mn$_3$X compounds (with X either Ge or Sn) stand out. They are formed in a geometrically frustrated structure where the Mn sites form a layered Kagom\'e lattice that is slightly distorted around the X atoms. This results in a chiral magnetic order, where equilateral triangles of Mn atoms have moments rotated 240$^\circ$ between them, as displayed in Fig.~\ref{elk-resultat}(a)  \cite{Zimmer:2008jp,Zimmer:2012kt,Shoichi:2013dw}.
Previous investigations of this class of materials is driven by the observation of an unusually large anomalous Hall effect (AHE) as well as a large spin Hall effect (SHE)~\cite{Kubler:2014eb,Zhang:2013hn,Suzuki:2017gj,Zhang:2017jb}. The former was until recently assumed only to exist in ferromagnets.
 These transport properties have been shown to be related to that the electronic structure has topological character close to the Fermi level with Weyl points and large Berry curvature \cite{Kuroda:2017eq,Li:2018hj}.

Curiously, Mn$_3$X posses a very small ferromagnetic moment of the order of a thousandth of a Bohr magneton (m$\mu_\mathrm{B}$)\cite{Zimmer:2012kt}. 
This experimentally observed weak ferromagnetism (wFM) has been confirmed by non-collinear electronic structure calculations~\cite{Sandratskii:1996gc}.
However, as was observed early on, in contrast to  traditional wFM, such as in $\alpha$-Fe$_2$O$_3$, the high symmetry of the crystal structure makes the standard symmetry arguments inapplicable.  While the so-called Dzyaloshinskii-Moriya (DM) pair interactions \cite{Dzyaloshinsky:1958br,Moriya:1960go}, 
are allowed by the $D_{6h}$ crystallographic point group symmetry, the triangular arrangement of the Mn sites ensures that there is no net torque on the Mn moments to deviate from a perfect N\'{e}el state, which would be necessary for an instability towards a configuration with a wFM component~\cite{Zimmer:2008jp,Shoichi:2013dw,Sandratskii:1996gc,
Sticht:1989dc}.


Effective spin Hamiltonians represent a common approach to understand the energies of different magnetic configurations of materials, and may be viewed as a mapping, from high energy electronic states to low energy spin-excitations. Since the pioneering works of Heisenberg, Dzyaloshinskii and Moriya, it has become standard practice to interpret magnetic phenomena by means of the Heisenberg Hamiltonian, augmented, when relativistic spin-orbit coupling (SOC) is important, with the Dzyaloshinskii-Moriya (DM) interaction and/or the anisotropic, symmetric exchange. The DM interaction has in its conventional form the property that it provides an energetic mechanism for the chirality of the magnetic state.

Hence, one of the more general considerations of a spin Hamiltonian is
\begin{align}
	\mathcal{H}&=-\sum_{ij}\left(J_{ij}\hat{m}_i{\cdot}\hat{m}_j+\vec{D}_{ij}{\cdot}(\hat{m}_i{\times}\hat{m}_j)+\hat{m}_i{\cdot}\mathcal{A}_{ij}{\cdot}\hat{m}_j\right), 
	\label{ham}
\end{align}
where $\hat{m}_i$ represent the unit vector of the magnetic moment at site $i$. The first term on the right hand side is the Heisenberg interaction, the second term is the DM interaction, while the last term is the anisotropic symmetric exchange interaction. Taken together they include all allowed bilinear spin-interactions.
There are several different ways one may calculate the parameters of Eq.~(\ref{ham}). Among the most popular methods is the approach to directly calculate them for a reference system, in a perturbational way.
Alternatively, one may fit these parameters to calculated variations in energies of different magnetic configurations. For systems with non-trivial, long range interaction (typically for metallic magnets) the exchange parameters are sensitive to the magnetic order. Therefore a fit to energies obtained for different magnetic configurations makes little sense. A preferable approach is then to calculate the parameters for a reference magnetic state, ideally the ground state. The most popular method of this type is the well established Liechtenstein-Katsnelson-Antropov-Gubanov approach~\cite{Liechtenstein:1987br}, that allows to calculate the Heisenberg interactions, $J_{ij}$. The pioneering work of Ref.~\cite{Liechtenstein:1987br} is valid in the limit of collinear reference states, and we note that there have been suggestions how to generalize this method for non-collinear magnetic reference states~\cite{Antropov:1997dra,Katsnelson:2000eg}. The DM interaction and the anisotropic exchange, i.e. the second and third term of the right hand side of Eq.~(\ref{ham}), are traditionally explained to be caused by the relativistic, spin-orbit interaction~\cite{Antropov:1997dra,Udvardi:2003ja,Ebert:2009cx}. Importantly, these two interactions are responsible for exotic magnetic phenomena, such as chiral magnetism and bond-directed, Kitaev exchange.


\section{The magnetic state of M\lowercase{n}$_3$S\lowercase{n}}
\begin{figure*}[t]
\begin{center}
\includegraphics[width=0.95\textwidth]{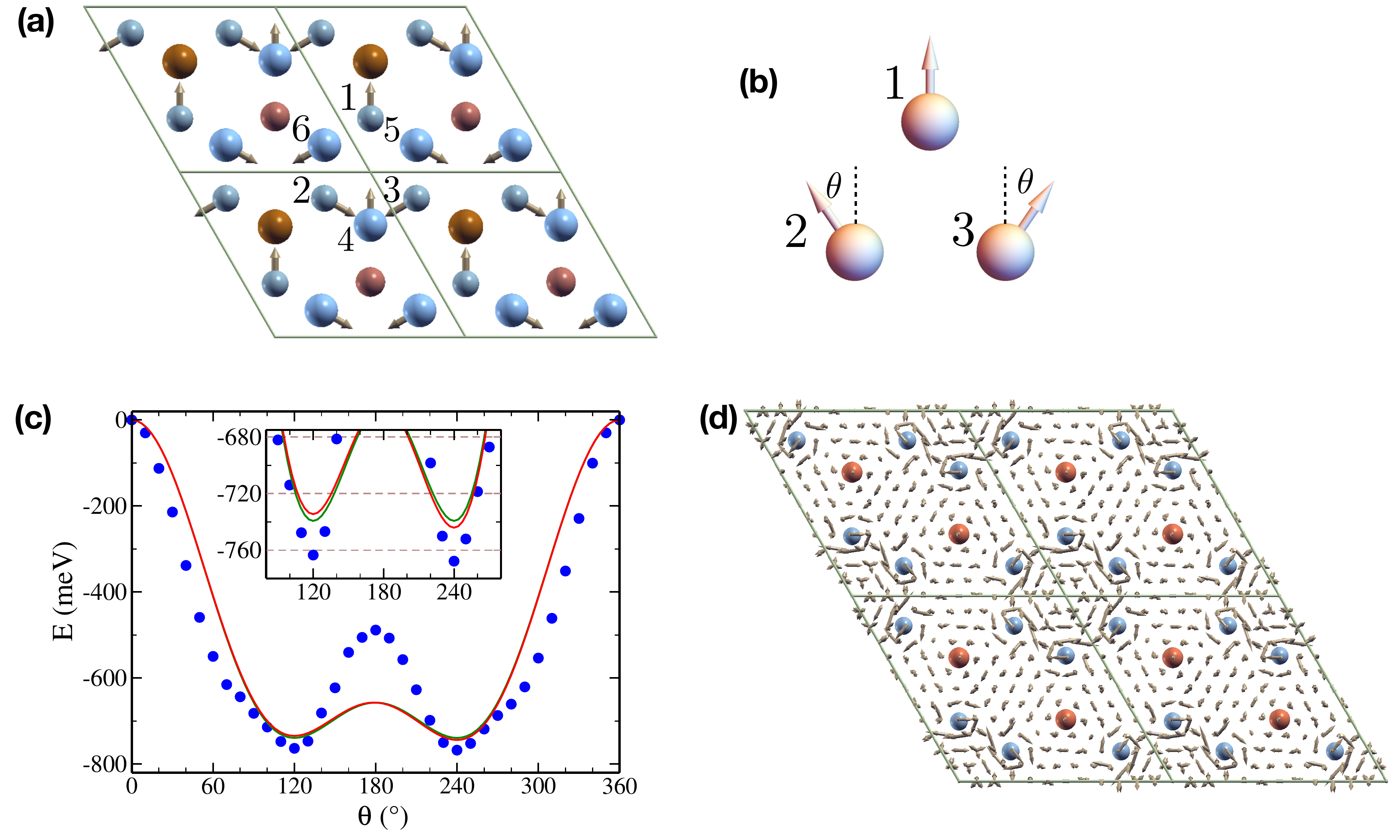}
\caption{
(a) The magnetic structure in case of negative vector chirality with $\theta=240^\circ$, illustrated for four unit cells of Mn$_3$X. There are two layers, with both Mn (blue) and Sn (red) atoms. In the lower layer, at $z=\frac{1}{4}$, atoms are shown as smaller and darker spheres, while upper layer, at $z=\frac{3}{4}$, have atoms shown as larger and lighter spheres. The lines highlight the unit cells, with six Mn atoms per cell, numbered from 1 to 6. Due to inversion symmetry the magnetic moments of atoms 4, 5 and 6  are parallel to the moments of 1, 2 and 3, respectively.  (b) Definition of the free angle, $\theta$, that can be varied in the magnetic structure of (a) illustrated for one Mn triangle. (c) The calculated total energy variation (blue points) within the $E_{1g}$ magnetic order and fits to Heisenberg interactions. In the inset the small difference when either the Dzyaloshinskii-Moriya is included in the fit (red curve) or not (green curve). (d) The spontaneous spin current polarized along $z$ for the magnetic structure depicted in (a). The shown spin currents were calculated considering the magnetic configuration when $\theta=240^{\circ}$.\label{elk-resultat}} 
\end{center}
\end{figure*}

Mn$_3$Sn belongs to the magnetic space group $\# 51.294$, ~\footnote{The notation of the corresponding space group follows the Belov-Neronova-Smirnova settings, see the link, www.cryst.ehu.es/cgi-bin/cryst/programs/nph-magtrgen?gnum=51.294, on the Bilbao Crystallographic Server.}, which corresponds to the two dimensional irreducible representation $E_{1g}$ ($\Gamma_5^+$) of the crystallographic point group $D_{6h}$ of the space group $\#194$. As is clear from Fig.~\ref{elk-resultat}(a), the Mn atoms are located on a Kagom\'e lattice, where the basic building block can be viewed as an ordering of equilateral triangles. The magnetic order of this material can be described as having local Mn moments within one triangle specified as follows, with $\theta$ angle defined in Fig.~\ref{elk-resultat}(b):
\begin{align}
		\hat{m}_1(\theta)&=\left(0,1,0\right)\nonumber\\
	\hat{m}_2(\theta)&=\left(-\sin\theta,\cos\theta,0\right)\nonumber\\
	\hat{m}_3(\theta)&=\left(\sin\theta,\cos\theta,0\right)\,.\label{magn_order}
\end{align}
It is well established that the ground state of Mn$_3$Sn corresponds to a state where the angle $\theta$ in Eq.~(\ref{magn_order}) is close to $240^\circ$ \cite{Zimmer:2008jp}. The results of accurate all-electron, full-potential augmented plane wave plus local orbitals calculation~\cite{Anonymous:2000js} of the energy variation within this non-collinear magnetic order is presented in Fig.~\ref{elk-resultat}(a). 
Before describing these results in more detail, we note that in terms of energy difference among the possible magnetic orders, the results from the electronic structure calculations are in good correspondence with earlier calculations~\cite{Sticht:1989dc,Sandratskii:1996gc}. From Fig.~\ref{elk-resultat}(c) it is clear that there is one minimum of the energy around $120^\circ$ and a second minimum at $240^\circ$. In the figure we also show the energy obtained from a fit of a spin model appropriate for the $E_{1g}$ symmetry, with two effective Heisenberg exchange parameters. Here the interactions are summed over the neighboring shell of atoms of given type; i.e. $J_{12}=\sum_{j\in 2}J_{1j}$
 and $J_{23}=\sum_{j\in 3}J_{2j}$. 
 The fitted curve in Fig.~\ref{elk-resultat}(c) was obtained using all the DFT calculated data points. Note that the fitted curve also leads to energy minima at $120^\circ$ and $240^\circ$, but with a much smoother energy variation, that does not capture the characteristic features of the energy variation obtained from the DFT calculations. The difference between the fitted curve and the DFT calculations becomes pronounced, most notably the cusps at $60^\circ$ and $300^\circ$ and the large energy barrier at $180^\circ$.
 
 A closer scrutiny of Fig.~\ref{elk-resultat}(c) reveals that the DFT calculations result in a minimum around $240^\circ$ which is lower by $4$ meV/u.c. compared to the minimum at $120^\circ$. In addition, the DFT calculations show that the absolute minimum is 
 shifted to angles that are slightly smaller than $240^\circ$.
 This results in a weak ferromagnetic component along the $\hat{y}$-direction, with a value $-0.002$ $\mu_\mathrm{B}$/u.c., which agrees qualitatively with experimental data. 
 We note that the observed wFM state cannot be explained by interactions that are connected to the fitted curve in Fig.~\ref{elk-resultat}(c), as the triangular symmetry of the Mn sites lead to DM interactions that perfectly balance out, so that
 a finite ferromagnetic moment of each unit cell is prohibited.

When discussing co-planar magnetic structures it is convenient to introduce a vector chirality of the magnetic order, which for the present order of Eq.~(\ref{magn_order}) is defined through the expression
\begin{align}
	\vec{\chi}(\theta)&=\vec{\chi}_{12}(\theta)+\vec{\chi}_{23}(\theta)+\vec{\chi}_{31}(\theta)=\chi(\theta)\,\hat{z}\,,\label{vec-chirality}
\end{align}
where $\vec{\chi}_{ij}(\theta)=\vec{m}_i\times\vec{m}_j$.
In this study where we focus on what effects are due to SOC and which are not (non-relativistic) we will utilize the extra symmetries that exist in the non-relativistic case since the spin space is decoupled from the real space.
Then it is possible to identify that without SOC there is always a symmetry corresponding to a uniform spin rotation of $180^\circ$ degrees around an axis in the plane of a co-planar magnetic order which swaps the chirality of the magnetic order. In our case with the choice of $y$-axis this symmetry leads to a degeneracy in energy between magnetic orders with $\theta$ and $(360^\circ-\theta)$, i.e.~between positive and negative vector chirality. 
It has been established that a non-vanishing vector chirality, $\vec{\chi}_{ij}$, is connected with a spontaneous spin-current between sites $i$ and $j$ \cite{Katsura:2005p9351}. To illustrate this, we show in Fig.~\ref{elk-resultat}(d) the $z$-polarized spin current of Mn$_3$Sn, in the case of $\theta=240^\circ$. The spin current was calculated through
\begin{align}
	\mathbf{q}_z(\mathbf{r})=\frac{1}{V_\mathrm{BZ}}\int_\mathrm{BZ} \Im\, \left\{\psi_\mathbf{k}^*(\mathbf{r})\sigma_z \boldsymbol{\nabla} \psi_\mathbf{k}^{\phantom{*}}(\mathbf{r})\right\}\,\mathrm{d}\mathbf{k}\,.
\end{align} 
In Fig.~\ref{elk-resultat}(d) it is clear that the spin-currents are particularly visible around the hexagonal shaped vortex structures. These structures are located around Mn triangles that are inverted every second layer, see Fig.~\ref{elk-resultat}(a). The largest spin currents are actually in between the layers forming compensated helices, all going around the vortex structure in a clockwise motion. 

\section{Mapping electronic energies on the spin Hamiltonian }
In this section, we consider the magnetic interactions, as described in Eq.~(\ref{ham}). We do this by a generalized form of the method by Liechtenstein {\em et al}.~\cite{Liechtenstein:1987br} (referred here to as LKAG form) so as to handle non-collinear magnetism. Starting from a non-collinear magnetic reference state $\{\hat{m}_i\}$, we study the variation in energy when the directions of the local moments of the crystal rotate as $\vec{m}_i'\rightarrow\vec{m}_i+\delta \vec{m}_i$, where $i$ denotes a magnetic site. It should be noted that, since we are interested in a non-collinear arrangement of magnetic moments, as was done in, e.g., Ref.~\cite {Katsnelson:2010fr}, it is sufficient to consider linear, single site variations of the magnetic moment. This is in contrast to the situation of collinear magnetism, where two site rotations are necessary~\cite{Liechtenstein:1987br}. We expand on this fact in Appendix A.
The variation in energy corresponding to a single site rotation of the magnetic moment is given through partial integration of the integrated band energies (via the magnetic force theorem), i.e.,
\begin{align}
	\delta E=\int\,\varepsilon\,\delta D\,\mathrm{d}\varepsilon=-\int\,\delta N\,\mathrm{d}\varepsilon\,.\label{deltaE}
\end{align}
The change in integrated density of states is given by Lloyd's formula
\begin{align}
	\delta N=-\frac{1}{\pi}\Im\, \mathrm{Tr}\, \ln \left( 1+\delta t\, G\right)=
	\frac{1}{\pi}\Im\, \mathrm{Tr}\, \sum_n	\, \frac{(-\delta t\, G)^n}{n}
	\,,\label{Lloyd}
\end{align}
where the change in scattering potential is due to the variation of the magnetic moment direction;
\begin{align}
\delta t=\sum_i \, {\Delta}_i \delta \vec{m}_i\, \cdot\vec{\sigma}\,.	
\label{dt}
\end{align}
In Eqs.~(\ref{Lloyd}) and (\ref{dt}), $\Delta_i$ is the spin dependent part of the potential (exchange splitting) of site $i$, $G$ is the intersite Green function (GF) and the trace, Tr, is over both the spin and orbital degrees of freedom. In a non-collinear state it is sufficient to study the linear term of Eq.~(\ref{Lloyd}), in contrast to the collinear limit where this contribution often vanishes, since then the variation of the local moment is orthogonal to all local moments. From this consideration we obtain
\begin{align}
\delta N\approx -\frac{1}{\pi}\Im\, \mathrm{Tr}\, \delta t\,G 
	=-\frac{2}{\pi}\Im\, \sum_i \,\delta \vec{m}_i\cdot\, \mathrm{tr}\,{\Delta}_i\,\vec{G}^0_{ii}\,,\label{1st-order}
\end{align}
where the last trace, tr, on the right hand side is only over the orbital sub-space. In the last step we have decomposed the GF into four parts \cite{Fransson:2017jz};
\begin{align}
	G_{ij}
	&=\left(G^{00}_{ij}+G^{01}_{ij}\right)\vec{1}+\left(\vec{G}^{0}_{ij}+\vec{G}^{1}_{ij}\right)\cdot\vec{\sigma}\,,\label{GF-st}
\end{align}
where $\vec{1}$ is the unit matrix while $\vec{\sigma}$ are the Pauli matrices.

In Eq.~(\ref{GF-st}), $G^{0\tau}$ corresponds to the spin-independent contributions to the GF, while $\vec{G}^\tau$ are the spin-dependent parts. The  superscript $\tau$ refers to whether the GF is even, 0, or odd, 1 under site exchange, which is readily introduced for a real basis. This is described in Appendix A together with a generalization to complex bases. There it is also demonstrated that one may relate the different components of the GF to physical properties, i.e.~charge and spin  densities, $n$ and $\vec{m}$,  as well as charge  and  spin currents, $\mathbf{j}$ and $\vec{\mathbf{q}}$.

Since the local GF can be evaluated within multiple-scattering theory, one can derive an explicit expression for it due to electron scattering of spin-potentials at any pair of sites, $i$ and $j$. As shown in Appendix A this leads to the following distinctly different contributions to the local, even part of the magnetic component of the GF:
\begin{align}
	\vec{G}^{0}_{ii}
	=&-\sum_j\left[
	G^{00}_{ij}\,\vec{\Delta}_j\,G^{00}_{ji}-G^{01}_{ji}\,\vec{\Delta}_j\,G^{01}_{ji}\right.
	\nonumber\\&
	 + \vec{G}^0_{ij}{\Delta}_j\cdot\vec{G}^0_{ji}\hat{m}_j-\vec{G}^1_{ij}{\Delta}_j\cdot\vec{G}^1_{ji}\hat{m}_j\nonumber\\&
	+ 2i\left(G^{00}_{ij}\,\vec{\Delta}_j\times\vec{G}^1_{ji}-G^{01}_{ij}\,\vec{\Delta}_j\times\vec{G}^0_{ji}\right)
	\nonumber\\&
+\vec{G}^1_{ij}\left(\vec{\Delta}_j\cdot\vec{G}^1_{ji}\right)+\left(\vec{G}^1_{ij}\cdot\vec{\Delta}_j\right)\vec{G}^1_{ji}\nonumber\\&
-\left.\vec{G}^0_{ij}\left(\vec{\Delta}_j\cdot\vec{G}^0_{ji}\right)-\left(\vec{G}^0_{ij}\cdot\vec{\Delta}_j\right)\vec{G}^0_{ji}
	\right]\,.\label{local-sum-rule}
\end{align}
This expression together with Eqs.~(\ref{deltaE})--(\ref{1st-order}), result in an explicit formula for the energy change due to rotating the moment at any lattice site. The terms that enter with linear order can with the aid of Eq.~(A.13) of Appendix A, be viewed as three different bilinear interactions, as shown in Eq.~(\ref{ham}), where the first terms is the isotropic Heisenberg interaction
\begin{align}
	J_{ij}=&\frac{1}{\pi}\Im\int \mathrm{tr}\Bigl(\Delta_i G^{00}_{ij}\Delta_jG^{00}_{ji}-\Delta_i G^{01}_{ij}\Delta_jG^{01}_{ji}
	\nonumber\\&
	+\Delta_i\vec{G}^{0}_{ij}\Delta_j\cdot\vec{G}^{0}_{ji}-\Delta_i \vec{G}^{1}_{ij}\Delta_j\cdot\vec{G}^{1}_{ji}\Bigr)\mathrm{d}\varepsilon.\label{defJ}
\end{align}
We note that in a non-collinear, coplanar magnetic configuration, all terms of linear order do not vanish, i.e.~$\delta\vec{m}_i\cdot\hat{m}_j$ is finite whenever $\delta\vec{m}_i$ is within the $xy$-plane of coplanar magnetic moments. According to the discussion above, we notice that four distinctly different contributions can be identified in the Heisenberg exchange interaction, which are related to a charge density (CD) contribution, a charge current (CC) contribution, a spin density (SD) component and a spin current (SC) term, respectively. Two of these contributions are unique for non-collinear systems, due to the fact there might be spontaneous spin and charge currents, when $\vec{G}^1$ and $G^{01}$ are non-vanishing (we remind the reader that a full account of the connection between different components of the Greens function and spin- and charge densities and currents is discussed in Appendix A). In this appendix it is shown that the relation in Eq.~(\ref{defJ}) reduces to the well known LKAG form of the Heisenberg interaction, in the non-relativistic collinear limit~\cite{Liechtenstein:1987br}. This is rewarding since the basic consideration made in the derivation of Eq.~(\ref{defJ}) was made from single site rotation out of a general non-collinear state, as opposed to 
the collinear limit used in Ref.~\onlinecite{Liechtenstein:1987br}.

Similar to the expression for Heisenberg exchange, one may obtain an explicit formula for the DM interaction:
\begin{align}
	\vec{D}_{ij}&=\frac{2}{\pi}\Re\int \mathrm{tr}\,\left(\Delta_i\, G^{00}_{ij}\,\Delta_j\,\vec{G}^{1}_{ji}-\Delta_i\, G^{01}_{ij}\,\Delta_j\,\vec{G}^{0}_{ji}\right)\,\mathrm{d}\varepsilon\,.\label{defD}
\end{align}
The DM interaction has two contributions; one arises from spin current $\vec{G}^1$ and the other from charge current $G^{01}$ (for details, see Appendix A). Since it is linear in current contributions, the anti-symmetric property falls out directly i.e.~that $\vec{D}_{ij}=-\vec{D}_{ji}$. The fact that DM is directly related to the SC induced by SOC have been observed by other means~\cite{Freimuth:2014en,Freimuth:2017ty,Kikuchi:2016gy}, but from this pair interaction formulation it is clear that there is also CC contribution and  
most noteworthy both these contributions can be non-zero, even when spin-orbit interaction is neglected, provided that a non-collinear arrangement of magnetic moments is considered. As we shall see, the non-relativistic contributions can in some cases dominate over terms that are associated with the relativistic spin-orbit coupling.

\section{On non-relativistic contributions to the Dzyaloshinskii-Moriya interaction }

\begin{figure*}
\begin{center}
\includegraphics[width=2\columnwidth]{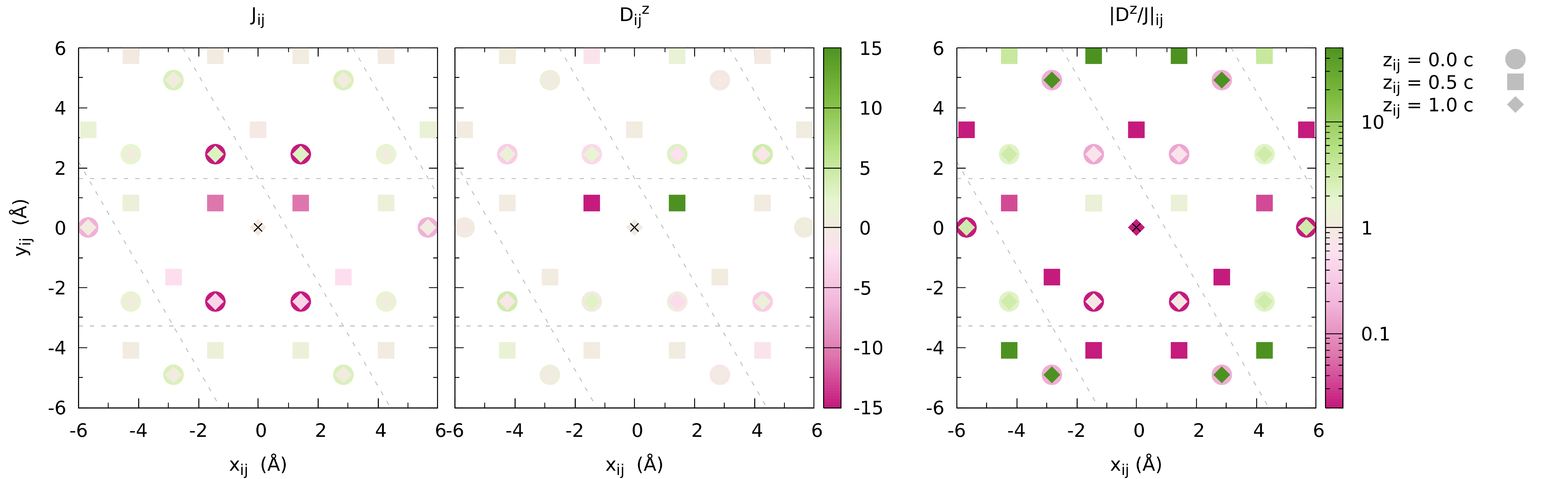}
\caption{The Heisenberg interaction (left figure), the Dzyaloshinskii-Moriya interaction (middle figure), as well as the ratio between these two interactions (right figure), of Mn$_3$Sn. The interactions are plotted as a function of distance for a magnetic configuration with $\theta=240^\circ$. The origin represents the position of the reference atom, $i$, and the interactions are taken between $i$ and its neighbors, $j$ with $\mathbf{r}_{ij}=(x_{ij},y_{ij},z_{ij})$. The value of $z_{ij}$, as indicated by the symbol shape, denotes in which atomic plane atom $j$ is situated, i.e. $z_{ij}=0.0c$ has atom $j$ in the same plane as atom $i$. Atom $j$ in the the first atomic plane in the $z$ direction corresponds to $z_{ij}=0.5c$, while $z_{ij}=1.0c$ means that $j$ is in the second atomic plane in this direction. Interactions are shown as function of distance along 
$x$-axis ($x_{ij}$) and 
$y$-axis ($y_{ij}$). The strength of the interactions are given by the color of the symbols. The two leftmost figures share the same color scale (in meV) where green is a strong positive interaction and purple strong negative interaction. The rightmost figure shows the ratio of the two interactions and has a logarithmic colorscale where green means the magnitude of the Dzyaloshinskii-Moriya interaction is larger than the corresponding Heisenberg interaction.}
\label{bubble}  
\end{center}
\end{figure*}

In order to analyze both the Heisenberg and DM interactions of Mn$_3$Sn, we show in Fig.~\ref{bubble} these interactions for a large range of pairs in the crystal, for the case when the magnetic configuration is described by $\theta=240^\circ$. These calculations were made with a real-space linear muffin-tin orbital (LMTO) method within the atomic sphere approximation~\cite{RS-LMTO92,RS-LMTO02,RS-LMTO06} (see Appendix A for details). It is clear from Fig.~\ref{bubble} that, although the exchange and DM interactions are long-ranged and oscillating, as is typical for metallic systems, they are dominated by the nearest neighbor terms. The nearest neighbour exchange interactions are all negative, and lead to frustrated antiferromagnetism.
In the same figure we also plot the $z$-component of the DM interaction. From Fig.~\ref{bubble} it is clear that the out of plane nearest neighbor interaction dominates the DM interaction, and, surprisingly, is of same magnitude as the Heisenberg interaction. In the right hand side of Fig.~\ref{bubble} we compare the strength of the two interactions, via the ratio $|D^z/J|_{ij}$. One may observe that for some atomic pairs, in particular for longer ranged interactions, the strength of the DM term is actually larger than the Heisenberg contribution. 

\begin{figure}
\begin{center}
\includegraphics[width=1\columnwidth]{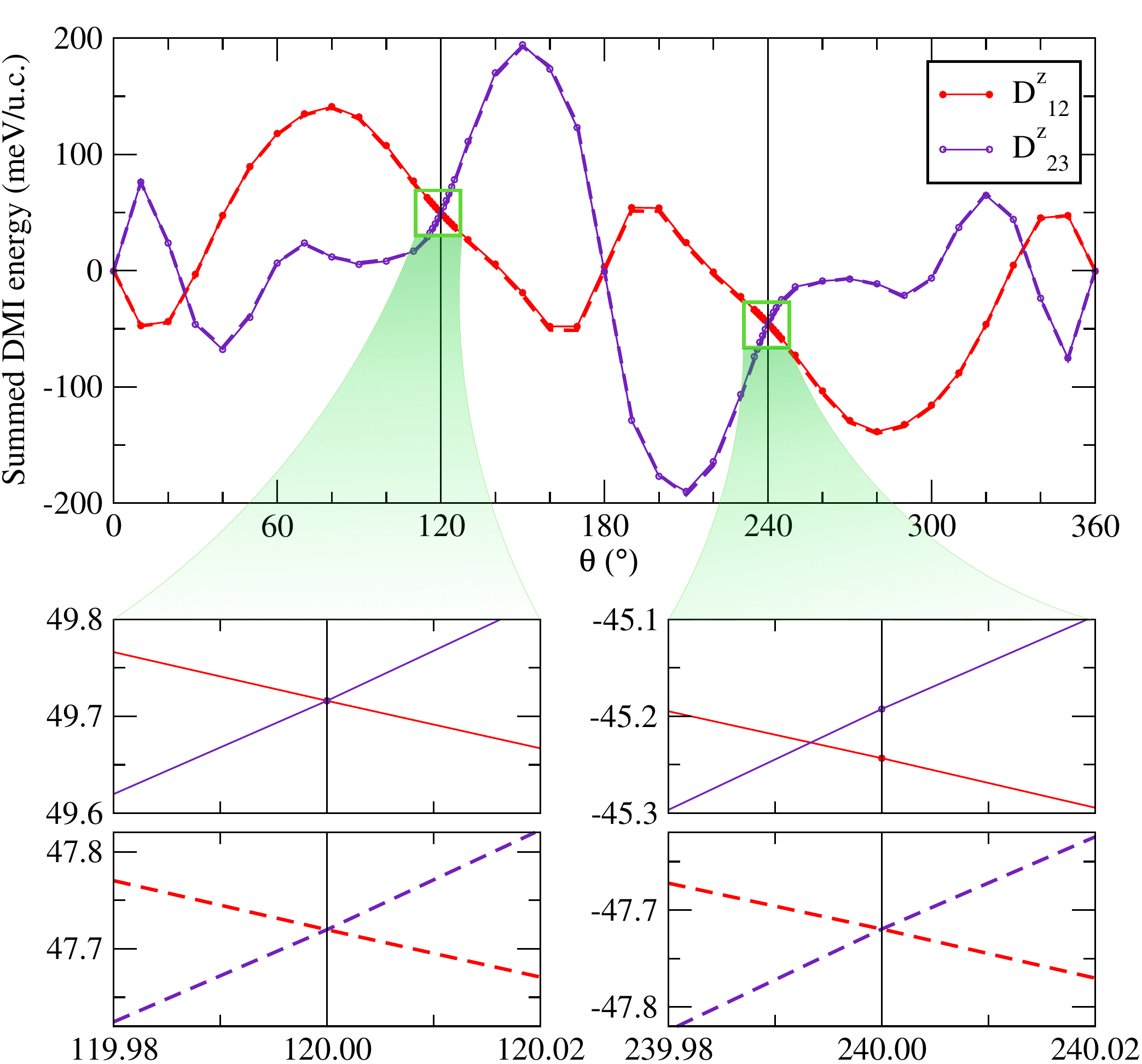}
\caption{Calculated results of the effective DM interactions $D^{z}_{12}$ and $D^{z}_{23}$ for Mn$_3$Sn, as a function of angle $\theta$ (for details see text). Calculations are made with (solid lines) and without (dashed lines) spin-orbit coupling. The bottom panels show detailed information around $\theta=120^\circ$ and $\theta=240^\circ$. \label{D-resultat}}  
\end{center}
\end{figure}

The configuration dependence of the DM interactions of Mn$_3$Sn is displayed in Fig.~\ref{D-resultat}. The figure illustrates the $\theta$-dependence of two effective DM interactions; $D^{z}_{12}=\sum_{j\in 2}D^{z}_{1j}$
 and $D^{z}_{23}=\sum_{j\in 3}D^{z}_{2j}$. Note that the variation of the angle $\theta$ is within the $E_{1g}$ sub-space, with the exception of $\theta=120^\circ$ which has higher symmetry. The figure shows results of two different calculations, one with and one without spin-orbit interaction, which are essentially indistinguishable. There are several important facts to observe in this figure, where the very large value of the effective DM interaction, even in absence of spin-orbit interaction, stands out. Calculations of the electronic structure in a non-collinear configuration can, according to Fig.~\ref{D-resultat}, have surprisingly large values of the DM vector, even when spin-orbit interaction is omitted. In fact, for Mn$_3$Sn we can see that the relativistic contribution to the DM vector is in general minor.
A deeper analysis of the configuration dependence of the DM interaction shows that it is the first term of Eq.~(\ref{defD}), which is connected to spin-currents, that dominates. This is due to the fact that co-planar magnetic structure in the present study allow for spontaneous non-relativistic SC but not CC.
Again this is due to a spin rotation symmetry, this time around an axis normal to the magnetic plane, together with the action of time reversal. The former action reverses all magnetic moment while the latter restores them. This combined symmetry disallows non-relativistic CC as well as the spin components of a SC in the magnetic plane. However the perpendicular spin component of a non-relativistic SC is always allowed in a co-planar magnet.
The idea that the DM originates from spin currents has been proposed in Refs.~\onlinecite{Kikuchi:2016gy,Freimuth:2017ty}, however, in both cases it was discussed in or close to the collinear limit, where the effect is of a pure relativistic origin. Spin currents induced from non-collinear spin-structures were only mentioned in passing \cite{Kikuchi:2016gy} and were not explicitly considered.  Lastly, it is noteworthy to mention that recently a new class of multi-spin interactions based on scalar and vector chiralities~\cite{Grytsiuk2020} has been proposed to emerge in non-collinear systems.

The non-relativistic contributions to the interactions, due to the symmetry discussed above, have the property that when the vector chirality is reversed, i.e.~going from $\theta$ to $360^\circ-\theta$, the sign of the DM interaction is changed, as is clear from Fig.~\ref{elk-resultat} and which is illustrated by the calculations without SOC in Fig.~\ref{D-resultat}. This is due to the fact that the non-relativistic DM is directly related to the non-relativistic spin current, which alter sign due to this non-relativistic spin symmetry. This has significant influence on the macroscopic magnetic properties of Mn$_3$Sn, as the configurations with  $120^\circ$ and $240^\circ$ is degenerate in the non-relativistic limit, both representing a ground state N\'{e}el configuration with vanishing macroscopic moment. Although the non-relativistic DM interaction has an identical form as the relativistic one, 
$\vec{D}\cdot\hat{m}_i\times\hat{m}_j$, i.e.~is explicitly proportional to the vector chirality $\chi_{ij}=\hat{m}_i\times\hat{m}_j$, it is more restricted due to symmetry constraints. However, when symmetry is allowed it most often dominates the relativistic counterpart.
For instance it does not give rise to a {\em chiral} state, a magnetic order with preferred {\em handedness}, in this case that the global vector chirality of Eq.~(\ref{vec-chirality}) has a preferred sign.
Instead the
 degeneracy discussed above 
 is broken by spin-orbit coupling, as shown in the bottom panels of Fig.~\ref{D-resultat}. In the bottom panels it may also be observed that for the configuration with $\theta = 240^\circ$, the effective pair interactions do not perfectly balance out, i.e. $D^{z}_{12} \ne D^{z}_{23}$. This leads to a ground state configuration that deviates from an ideal N\'{e}el state, with an angle $239.99^\circ$. This corresponds to a very small wFM moment of $-0.002\, \mu_\mathrm{B}/f.u.$ along the $y$-direction, in good agreement with experiments~\cite{Zimmer:2008jp}. 
Here it is important to realize that, except for angle $120^\circ$, the crystallographic point group corresponding to the magnetic space group is orthorhombic, $D_{2h}$, rather than hexagonal, $D_{6h}$. When it is accepted that the DM interactions are configuration dependent there is no longer a symmetry equivalence between $D_{12}$ and $D_{23}$ for $240^\circ$ in the relativistic case, while for $120^\circ$, where the hexagonal symmetry gives an equivalence, the splitting is indeed zero.
We further observe that the splitting of $D^{z}_{12}$ and $D^{z}_{23}$ is zero at both angles in calculations without spin-orbit coupling, since we then have hexagonal symmetry in both cases.

Our findings are in contrast to what has earlier been stated \cite{Zimmer:2012kt,Sandratskii:1996gc}; that the DM interaction is irrelevant for the wFM of \mx. The mechanism behind our results also differ from the traditional mechanism in numerous other materials, such as $\alpha-\mathrm{Fe}_2\mathrm{O}_3$, where the the DM interaction has successfully been invoked \cite{Dzyaloshinsky:1958br,Moriya:1960go} to explain the weak ferromagnetic moment observed experimentally. Let us identify what differ between our finding and the reasoning why DM interaction cannot be responsible.
\begin{enumerate}[label=(\roman*)]
    \item The DM parameter $\vec{D}_{ij}$ is not independent of magnetic configuration, as is usually assumed. Instead it is found to vary between different magnetically ordered states.
    \item  The DM interaction does not have to fulfill the symmetries of the paramagnetic state, but only those of the magnetic state. For instance for the case of the $240^\circ$ state there exist no three-fold rotations, since the symmetry group in fact has an orthorhombic magnetic space group, which means there is no relation between the two independent DM parameters.
    \item Hence, there is a splitting in value at $240^\circ$ between the two independent DM interactions, labelled $D_{12}^z$ and $D_{23}^z$, as they are not forced by symmetry to be equivalent and therefore will not balance each other.
    \item The balance between the DM parameters occur at a new angle that deviates from $240^\circ$.
    For such an angle the anti-ferromagnetic moments do not cancel which results in a weak ferromagnetic moment. 
    \item
    The splitting of the two DM parameters have to be of relativistic origin due to the the non-relativistic spin rotation symmetry makes the non-relativistic part of DM to posses the three-fold rotational symmetry of the $120^\circ$ case. 
    \item 
    However,
    as observed in the upper rightmost panel of Fig.~\ref{D-resultat}, the splitting $\Delta D(\theta)$ is well described as being linear in angle, when deviating from $\theta_0=240^\circ$, i.e. 
    \begin{align}
        \Delta D(\theta)= \Delta D(\theta_0)+(\theta-\theta_0) K\,,
    \end{align}
    with slpoe $K$.
    The new equilibrium angle $\theta$ for which $\Delta D(\theta)=0$ is given by
    \begin{align}
        \theta=\theta_0-\frac{\Delta D(\theta_0)}{K}\,.
    \end{align}
    This resulting angle and corresponding wFM moment is in good accordance with experimental values.
    \item It is mainly the non-relativistic contribution that contribute to the slope $K$. When only relativistic part of DM is considered the resulting wFM moment would be
    orders of magnitude larger.
\end{enumerate}


\section{Conclusions}

In this report, we demonstrated that magnetic interactions can in general be mediated in four different ways, that are related to the charge density, spin density, charge current and spin current, respectively, and that they do  depend strongly on the magnetic configuration. This is demonstrated using Mn$_3$Sn as an example, where we present in the main part of this communication a detailed analysis of the Dzyaloshinskii-Moriya interaction. A full account of all three types of exchange interactions defined in Eq.~(\ref{ham}) is presented in Appendix B, where in particular the configuration dependence of the Heisenberg exchange is analyzed in detail.
The spin-current contribution to the Dzyaloshinskii-Moriya interaction is shown here to be significant, and it is particularly relevant for the co-planar magnetic material Mn$_3$Sn. Spin-currents induced by non-collinear states are related to the large spin- and anomalous Hall effects observed for Mn$_3$Sn~\cite{Kubler:2014eb,Zhang:2013hn,Suzuki:2017gj,Zhang:2017jb}.
Our analysis points to that the very small ferromagnetic component of the magnetism of Mn$_3$Sn is 
due to a combined effect of the relativistic and the non-relativistic DM interactions. 

Most significantly, the Dzyaloshinskii-Moriya interaction is demonstrated here to be very large for Mn$_3$Sn; for some pair interactions its magnitude is comparable to the Heisenberg exchange. The analysis presented here, which has general applicability, demonstrates that the Dzyaloshinskii-Moriya interaction (as well as the anisotropic, symmetric exchange) can be significant even if spin-orbit interaction is omitted. In fact, for Mn$_3$Sn non-relativistic effects dominate. However it does not, due to symmetry constrains, contribute to the preferred {\em chiral handedness} of the ground state, which is determined solely by the relativistic contribution for any co-planar magnetic structure.

The microscopic reason behind the non-relativistic contribution to the DM interaction is analyzed here to be caused by the contribution from spin-currents, that are set in motion either by spin-orbit coupling or by non-collinear magnetic configurations. The reason why these two, seemingly different, microscopic mechanisms both result in spin-currents can be traced back to the electronic Hamiltonian. In Ref.~\cite{PhysRevLett.76.4420}, a mathematical similarity was established between the spin-orbit interaction and the exchange-correlation potential of non-collinear magnets. If the non-collinear contribution to the exchange and correlation potential is larger than the strength of the spin-orbit coupling, as is the case for Mn$_3$Sn, the former contribution dominates. As discussed in Appendix B, this realization suggest that in materials with non-collinear magnetism, all effects normally associated with spin-orbit coupling, should, due to induced spin-currents, have a counterpart in the non-collinear exchange and correlation potential.

\section{Acknowledgements}
We acknowledge financial support from CAPES and CNPq, Brazil, as well as from the Swedish Research Council (VR), Knut and Alice Wallenberg Foundation (KAW), Energimyndigheten, the Foundation for Strategic Research (SSF), STandUPP and eSSENCE, Sweden.
The calculations were performed at
the computational facilities of the CENAPAD at University of Campinas, SP, Brazil
and NSC in Link\"oping under allocations provided by SNIC.

\section{Appendix A: Calculation details}
        \setcounter{table}{0}
        \renewcommand{\thetable}{A\arabic{table}}%
        \setcounter{figure}{0}
        \renewcommand{\thefigure}{A\arabic{figure}}%
        \renewcommand{\theequation}{A.\arabic{equation}}%
        \setcounter{equation}{0}

\subsection{Computational details}
Full potential electronic structure calculations were performed using the ELK package (\textit{elk.sourceforge.net}). This represents a set of accurate, non-collinear, full-potential augmented plane wave plus local orbitals calculations~\cite{Anonymous:2000js}. In these calculations we used $R_\mathrm{Mn}G_\mathrm{max}= 9$ and a k-point mesh of $24\times24\times20$. Furthermore, the Perdew-Wang local density functional~\cite{Perdew:1992ee} for exchange and correlation was used. The experimental crystal structure of Mn$_{3}$Sn was considered in the calculations. Where applicable, spin-orbit coupling was treated at the variational step.

The real-space electronic structure calculations~\cite{RS-LMTO92,RS-LMTO02,RS-LMTO06} were performed within the local spin density approximation for the exchange correlation energy of density functional theory. When applicable, the spin-orbit coupling was included at each variational step. The real-space calculations employed linear muffin-tin orbitals (LMTO) as basis functions~\cite{LMTO}. Furthermore, the atomic sphere approximation  was adopted, and the computational method is here referred to as the RS-LMTO-ASA method. 
These self-consistent, non-collinear calculations employed Haydocks recursion method \cite{HAYDOCK1980215}.
The Mn$_{3}$Sn crystal was simulated by a cluster containing 17.000 atoms located in a lattice where the Mn sites form a layered Kagom\'e lattice that is slightly distorted around the Sn atoms (see Fig.~\ref{elk-resultat}(a). The continued fraction that occurs in the recursion method was terminated with the Beer-Pettifor~\cite{Beer1984} terminator after 30 recursion levels. The values of $J_{ij}$, $D_{ij}$ and $A_{ij}$ of Eq.~(\ref{ham}) were obtained from a self-consistent non-collinear calculations performed with and without spin-orbit coupling, for magnetic structures with triangles of Mn atoms having the moments rotated from $\theta=0^\circ$ until $360^\circ$, with the $\theta$ angle defined in Fig.~\ref{elk-resultat}(b), and varying every $10^\circ$.

\subsection{On notations and formalism}

The multiple scattering condition for the local Green's function (GF) can be formulated as 
\begin{align}
	\left(G^{-1}-G_0^{-1}\right)_{ij}=\left(T_j+t_j^\mathrm{soc}\right)\delta_{ij}\,,\label{MS} 
\end{align}
where $G_{0;ij}$ is the free GF, which is spin-independent, and $T_j=t_j+\hat{m}_j\Delta_j\cdot\sigma$, where $t_j$ stands for the local spin-independent scattering potential. All quantities are matrices in a combined spin and orbital basis.
We now express the GF in terms of spin-polarized scattering theory, for a non-collinear ordered reference state and including the spin-orbit coupling (SOC). Therefore, we introduce 
the GF for the time reversed reference state, $\tilde{G}_{ij}$, i.e.~for the case with all moments reversed, together with the directions of charge current. We obtain;
\begin{align}
	\left(\tilde{G}^{-1}-G_0^{-1}\right)_{ij}=\left(\tilde{T}_j+t_j^\mathrm{soc}\right)\delta_{ij}\,.\label{MS-TR} 
\end{align}
Both $G_0$ and $t_i^\mathrm{soc}$ are even under time reversal (TR), while $\tilde{T}$ is the time reversed scattering potential.
Now the difference of Eqs.~(\ref{MS}) and
(\ref{MS-TR}) gives
\begin{align}
	\left({G}^{-1}-\tilde{G}^{-1}\right)_{ij}=\left(T_j-\tilde{T}_j\right)\delta_{ij}=2\,\hat{m}_j\Delta_j\cdot\sigma\,\delta_{ij}\,,\label{MS-diff}
\end{align}
since $\tilde{T}_j=t_j-\hat{m}_j{\Delta}_j\cdot\sigma$.
By letting $\tilde{G}$ and $G$ act on Eq.~(\ref{MS-diff}) from either side, we get for the local GF
\begin{align}
	{G}_{ii}-\tilde{G}_{ii}=-\,\sum_j \hat{m}_j\left(G_{ij}\,{\Delta}_j\cdot\sigma\,\tilde{G}_{ji}+\tilde{G}_{ij}\,{\Delta}_j\cdot\sigma\,{G}_{ji}\right)\,.\label{TR-diff}
\end{align}
It is useful to decompose the corresponding $2\times 2$ real space Green's function, $G(\mathbf{r},\mathbf{r}';\omega)$, into four independent contributions
\begin{align}
	G(\mathbf{r},\mathbf{r}';\omega)&=\sum_{\nu\in\{0,x,y,z\}}\sum_{\mu=0}^1 \sigma_\nu\,G^{\nu\mu}(\mathbf{r},\mathbf{r}';\omega)\,,
\end{align}
where $\sigma_0$ is the identity matrix, while $\vec{\sigma}=\{\sigma_x,\sigma_y,\sigma_z\}$ are the Pauli matrices. 
In this expression, $\nu=0$ corresponds to the non-magnetic component of the Green's function, while $\nu=x,y$ or $z$ represents spin-polarized components, that we for brevity write in vector form; $\vec{G}=\{G^{x},G^{y},G^{\omega}\}$.
The second index, $\mu$, of the Green's function in Eq. (A.5), indicates whether the function is even (0) or odd (1) under the exchange of spatial coordinates ($\mathbf{r}\leftrightarrow \mathbf{r}'$) and we write 
\begin{align}
	G^{\nu\mu}(\mathbf{r}',\mathbf{r};\omega)=(-1)^\mu\,G^{\nu\mu}(\mathbf{r},\mathbf{r}';\omega)\,.
\end{align}
Hence, we denote in general the spin-dependent GF as vectors $\vec{G}^\mu=\{G^{x\mu},G^{y\mu},G^{z\mu}\}$.
The four different Green's function discussed above all have a direct physical property as in the local limit they give rise to charge and spin density and charge and spin currents, respectively, through
\begin{align}
	n(\mathbf{r})=&-\frac{1}{\pi}\Im \int G^{00}(\mathbf{r},\mathbf{r};\omega)\,\mathrm{d}\omega\label{n-CD}\\
	\vec{m}(\mathbf{r})=&-\frac{1}{\pi}\Im \int \vec{G}^{0}(\mathbf{r},\mathbf{r};\omega)\,\mathrm{d}\omega\label{m-SD}\\
	\mathbf{j}(\mathbf{r})=&-\frac{1}{\pi}\Re \int \boldmath{\nabla} G^{01}(\mathbf{r},\mathbf{r};\omega)\,\mathrm{d}\omega\label{j-CC}\\
	\vec{\mathbf{q}}(\mathbf{r})=&-\frac{1}{\pi}\Re \int \boldmath{\nabla} \vec{G}^{1}(\mathbf{r},\mathbf{r};\omega)\,\mathrm{d}\omega\label{q-SC}\,.
\end{align}
As a side note, we observe that this formulation for the spontaneous currents in magnetic system directly ensures that they are source free.

With the discussion above in mind, it becomes relevant to decompose the GF matrix of the multiple scattering problem into four parts (assuming a real orbital basis);
\begin{align}
	G&=\left(G^{00}+G^{01}\right)\vec{1}+\left(\vec{G}^{1}+
	\vec{G}^{0}\right)\cdot\vec{\sigma}\,.\label{GF}
\end{align}
This decomposition will be performed in detail below where the case of complex bases is also discussed.
It is easy to deduce that two of the components of the Green's function in Eq. (A.11) are odd under time reversal, $G^{01}$ and $\vec{G}^0$, while the other two are even.
This implies that for the time-reversed state we have
\begin{align}
	\tilde{G}
	&=\left(G^{00}-G^{01}\right)\vec{1}+\left(\vec{G}^{1}-\vec{G}^{0}\right)\cdot\vec{\sigma}\,.\label{tauGF-st}
\end{align}
By utilizing Pauli spin matrix algebra, Eqs.~(\ref{TR-diff}), (\ref{GF}) and (\ref{tauGF-st}) lead to that for the TR odd spin dependent part $\vec{G}^0$
we have with $\vec{\Delta}_j=\hat{m}_j\Delta_j$
\begin{align}
	-\vec{G}^0_{ii}=\sum_j&\left[
	G^{00}_{ij}\,\vec{\Delta}_j\,G^{00}_{ji}-G^{01}_{ji}\,\vec{\Delta}_j\,G^{01}_{ji}+\right.\nonumber\\
	&
	+\left.\vec{G}^1_{ij}\left(\vec{\Delta}_j\cdot\vec{G}^1_{ji}\right)-\vec{G}^0_{ij}\left(\vec{\Delta}_j\cdot\vec{G}^0_{ji}\right)+\right.\nonumber\\
	&
	+2i\left(G^{00}_{ij}\,\vec{\Delta}_j\times\vec{G}^1_{ji}-G^{01}_{ij}\,\vec{\Delta}_j\times\vec{G}^0_{ji}\right)
	+\nonumber\\
&+\left.\vec{G}^0_{ij}\times\left(\vec{\Delta}_j\times\vec{G}^0_{ji}\right)-\vec{G}^1_{ij}\times\left(\vec{\Delta}_j\times\vec{G}^1_{ji}\right)
	\right]=\nonumber\\
	=\sum_j&\left[
	G^{00}_{ij}\,\vec{\Delta}_j\,G^{00}_{ji}-G^{01}_{ji}\,\vec{\Delta}_j\,G^{01}_{ji}+\right.\nonumber\\
	&+\left.\vec{G}^0_{ij}{\Delta}_j\cdot\vec{G}^0_{ji}\hat{m}_j-\vec{G}^1_{ij}{\Delta}_j\cdot\vec{G}^1_{ji}\hat{m}_j+\right.\nonumber\\
	&+2i\left(G^{00}_{ij}\,\vec{\Delta}_j\times\vec{G}^1_{ji}-G^{01}_{ij}\,\vec{\Delta}_j\times\vec{G}^0_{ji}\right)
	+\nonumber\\
&+\left.\vec{G}^1_{ij}\left(\vec{\Delta}_j\cdot\vec{G}^1_{ji}\right)+\left(\vec{G}^1_{ij}\cdot\vec{\Delta}_j\right)\vec{G}^1_{ji}+\right.\nonumber\\
&-\left.\vec{G}^0_{ij}\left(\vec{\Delta}_j\cdot\vec{G}^0_{ji}\right)-\left(\vec{G}^0_{ij}\cdot\vec{\Delta}_j\right)\vec{G}^0_{ji}
	\right]\,.\label{local-TR-odd}
\end{align}
It is the right hand side of Eq.~(\ref{local-TR-odd}) that should go into Eq.~(8) of the main text as $\delta N_1$ has to be TR even while $\Delta_i$ is TR odd.
This expression reduces to a well-known relations in the non-relativistic and collinear limit \cite{Liechtenstein:1987br,Antropov:1997dra,Katsnelson:2000eg},  where both $G^{01}$ and $\vec{G}^1$ vanish. With the moments along $\hat{z}$ we have that $\vec{G}^0=\frac{1}{2}(G_\uparrow-G_\downarrow)\hat{z}$ and hence
\begin{align}
	-\vec{G}^0_{ii}&=\hat{z}\sum_j{G_\uparrow}_{ij}\Delta_j{G_\downarrow}_{ji}\,.
\end{align}

\subsection{Complex basis}
To analyse differences in the formalism that arise to choice of basis (real or complex), we consider again the expression of the GF, i.e. 
\begin{align}
	G(\mathbf{r},\mathbf{r}';\omega)=\sum_{\eta}\,\sigma_\eta\, {G}^\eta(\mathbf{r},\mathbf{r}';\omega)\,,
\end{align}
with $\eta\in\{0,x,y,z\}$ and $\sigma_0={\bar 1}$ (note that $\nu$ in Eq. (A.5) now has been replaced by $\eta$). Then the decomposed GF is given  by
\begin{align}
	{G}^\eta(\mathbf{r},\mathbf{r}';\omega)=\frac{1}{2}\mathrm{Sp}\,G(\mathbf{r},\mathbf{r}';\omega)\,\sigma_\eta=\frac{1}{2}\sum_n\mathrm{tr}\,\frac{\sigma_\eta\langle{\mathbf{r}}|{n}\rangle\langle{n}|{\mathbf{r}'}\rangle}{\omega-\varepsilon_n}\,,
\end{align}
where $\mathrm{Sp}$ is the trace over the spin degree of freedom and $\mathrm{tr}$ is the trace over the orbitals. 
Next, we expand the GF,
 in a basis set $|{\nu}\rangle$,
\begin{align}
	G(\mathbf{r},\mathbf{r}';\omega)&=\langle{\mathbf{r}}|{n}\rangle\,(\omega-\varepsilon_n)^{-1}\,\langle{n}|{\mathbf{r}'}\rangle=\nonumber\\
	&=\langle{\mathbf{r}}|{\nu}\rangle\langle{\nu }|{n}\rangle\,(\omega-\varepsilon_n)^{-1}\,\langle{n}|{\nu'}\rangle\langle{\nu'}|{\mathbf{r}'}\rangle\nonumber\\
	&\equiv \langle{\mathbf{r}}|{\nu}\rangle \, G_{\nu\nu';\sigma\sigma'} \,\langle{\nu'}|{\mathbf{r}'}\rangle\,.
\end{align}
The spin decomposed GF then becomes
\begin{align}
	G^\eta(\mathbf{r},\mathbf{r}';\omega)&=\frac{1}{2}\mathrm{Sp}\,G(\mathbf{r},\mathbf{r}';\omega)\,\sigma_\eta=\nonumber\\
	&=\frac{1}{2}\langle{\mathbf{r}}|{\nu}\rangle\,\langle{\nu'}|{\mathbf{r}'}\rangle \, \mathrm{Sp}\, G_{\nu\nu';\sigma\sigma'}\,\sigma_\eta =\nonumber\\
	&\equiv \langle{\mathbf{r}}|{\nu}\rangle\,\langle{\nu'}|{\mathbf{r}'}\rangle \, G^\eta_{\nu\nu'}
\end{align}
while, in the same way,
\begin{align}
	G^\eta(\mathbf{r}',\mathbf{r};\omega)&=\langle{\mathbf{r}'}|{\nu'}\rangle \, G^\eta_{\nu'\nu} \,\langle{\nu}|{\mathbf{r}}\rangle\,.
\end{align}

It is possible to decompose the GF of Eq. (A.19), depending how it behaves under interchange of $\mathbf{r}\rightarrow \mathbf{r}'$. First we decompose the spin decomposed GF in a symmetric $G^{\eta0}$ and anti-symmetric part $G^{\eta1}$ 
with respect to the inter-exchange of $\mathbf{r}$ and $\mathbf{r}'$, i.e.~$G^\eta(\mathbf{r},\mathbf{r}';\omega)=G^{\eta0}(\mathbf{r},\mathbf{r}';\omega)+G^{\eta1}(\mathbf{r},\mathbf{r}';\omega)$ . 
We get with $\tau\in\{0,1\}$
\begin{align}
	G^{\eta\tau}(\mathbf{r},\mathbf{r}';\omega)&\equiv\frac{1}{2}\left\{G^\eta(\mathbf{r},\mathbf{r}';\omega)+(-1)^\tau\, G^\eta(\mathbf{r}',\mathbf{r};\omega)\right\}=\nonumber\\
	&=\frac{1}{2}\left\{\langle{\mathbf{r}}|{\nu}\rangle \, G^\eta_{\nu\nu'} \,\langle{\nu'}|{\mathbf{r}'}\rangle+(-1)^\tau \langle{\mathbf{r}'}|{\nu'}\rangle \, G^\eta_{\nu'\nu} \,\langle{\nu}|{\mathbf{r}}\rangle\right\}
	\,.\label{eq:pm2point}
\end{align}
If $\langle{\mathbf{r}}|{\nu}\rangle=\langle{\nu}|{\mathbf{r}}\rangle$, i.e.~the basis $\phi_\nu(\mathbf{r})=\langle{\mathbf{r}}|{\nu}\rangle$ is real, this can be written as
\begin{align}
	G^{\eta\tau}(\mathbf{r},\mathbf{r}';\omega)&=\frac{1}{2}\,\langle{\mathbf{r}}|{\nu}\rangle \left\{ G^\eta_{\nu\nu'} +(-1)^\tau  \, G^\eta_{\nu'\nu} \right\}\langle{\nu'}|{\mathbf{r}'}\rangle=\nonumber\\
	&=\frac{1}{2}\langle{\mathbf{r}}|{\nu}\rangle\left\{G^\eta +(-1)^\tau {G^\eta}^{\,t}\right\}_{\nu\nu'}\,\langle{\nu'}|{\mathbf{r}'}\rangle .
\end{align}
With a complex basis $\langle{\mathbf{r}}|{\nu}\rangle=\langle{\nu}|{\mathbf{r}}\rangle^*$ (e.g. a basis based on spherical harmonics, $|{\nu}\rangle=|{i\ell m}\rangle$), the analysis becomes a little more complicated. 
In this case, Eq.~(\ref{eq:pm2point}) takes the form 
\begin{align}
	G^{\eta\tau}(\mathbf{r},\mathbf{r}';\omega)&=\frac{1}{2}\left\{\langle{\mathbf{r}}|{\nu}\rangle \, G^\eta_{\nu\nu'} \,\langle{\nu'}|{\mathbf{r}'}\rangle\right.\nonumber\\
	&\left.+(-1)^\tau \langle{\mathbf{r}'}|{\nu'}\rangle \, G^\eta_{\nu'\nu} \,\langle{\nu}|{\mathbf{r}}\rangle\right\}=\nonumber\\
	&=\frac{1}{2}\left\{\langle{\mathbf{r}}|{\nu}\rangle \, G^\eta_{\nu\nu'} \,\langle{\nu'}|{\mathbf{r}'}\rangle\right.\nonumber\\
	&\left.+(-1)^\tau \langle{\nu'}|{\mathbf{r}'}\rangle^* \, G^\eta_{\nu'\nu} \,\langle{\mathbf{r}}|{\nu}\rangle^*\right\}
	=\nonumber\\
	&=\langle{\mathbf{r}}|{\nu}\rangle\,G^{\eta\tau}_{\nu\nu'}\,\langle{\nu'}|{\mathbf{r}'}\rangle
	\,.\label{G-cmplx}
	\end{align}
	The last step is possible if we can get a relation between $\langle{\mathbf{r}}|{\nu}\rangle^*$ and $\langle{\mathbf{r}}|{\nu}\rangle$ and if they both are in the basis, so that
\begin{align}
	G^{\eta\tau}_{\nu\nu'}&= \frac{1}{2}\left\{G^\eta_{\nu\nu'}+(-1)^\tau U_{\nu'\nu''} G^\eta_{\nu''\nu'''}U_{\nu'''\nu}^*\right\}\,.	
\end{align}	
In this expression, $U$ is the basis transformation that bring $\langle{\nu}|$ to $\langle{\nu'}|\equiv\langle{\nu}|^*=U_{\nu'\nu}\langle{\nu}|$.
	For spherical harmonics $\langle{\mathbf{r}}|{\nu}\rangle^*=\langle{\mathbf{r}}|{i\ell m}\rangle^*=Y_{\ell m}^*(\mathbf{r}-\mathbf{R}_i)=(-1)^m\,Y_{\ell -m}(\mathbf{r}-\mathbf{R}_i)=(-1)^m\langle{i\ell -m}|{\mathbf{r}}\rangle$ or $U_{\ell m';\ell m}=(-1)^m\delta_{m',-m}$.
	
	So for a GF defined in a spherical harmonics basis we have that
\begin{align}
	G^{\eta\tau}_{\nu\nu'}&= \frac{1}{2}\left\{G^\eta_{\nu\nu'}+(-1)^\tau U_{\nu'\nu''} G^\eta_{\nu''\nu'''}U_{\nu'''\nu}^*\right\}=\nonumber\\
	&=\frac{1}{2} \left\{ G^\eta_{i\ell m;j\ell' m'} +(-1)^{\tau+m+m'}\, G^\eta_{j\ell' -m';i\ell -m} \right\}=\nonumber\\
	&=\frac{1}{2}\left[ \{G^\eta_{ij}\}_{\ell m;\ell' m'} +(-1)^{\tau+m+m'}\, \{G^\eta_{ji}\}_{\ell' -m';\ell -m} \right]\,.
	\end{align}

\section{Appendix B: Configuration dependence}
        \setcounter{table}{0}
        \renewcommand{\thetable}{B\arabic{table}}%
        \setcounter{figure}{0}
        \renewcommand{\thefigure}{B\arabic{figure}}%
        \renewcommand{\theequation}{B.\arabic{equation}}%
        \setcounter{equation}{0}
The spin-configuration dependence (i.e. the $\theta$-dependence) of the Heisenberg interactions are shown in Fig.~\ref{J12-resultat}, for the effective strength of the interactions between atom 1 and equivalent atoms 2, as well as between atom 2 and equivalents atoms 3 (as defined in the main text). The figure has 
interactions decomposed in the four terms described in Eq.~(11) of the main text. It is clear from the figure that the charge-current (CC) term is essentially zero for all values of $\theta$. This is directly connected to the fact that the spontaneous CC is only due to SOC in the co-planar variation. This term is however expected to be important in a general, non-collinear structure. The charge-density (CD) term shows a rather weak $\theta$-dependence. In contrast, the spin-density (SD) and spin-current (SC) contributions are seen to depend strongly on the magnetic structure. This reflects that the magnetic configuration has large influence on the spin currents.
Comparing the SD and SC terms, we note that the former is fairly weak when the magnetic moments are far from a ferromagnetic configuration, but dominates in the ferromagnetic case. Lastly, we note from the figure that the spin current mediated terms are zero in the collinear limit, but they can be quite significant away from it.

\begin{figure}
\begin{center}
\includegraphics[width=1\columnwidth]{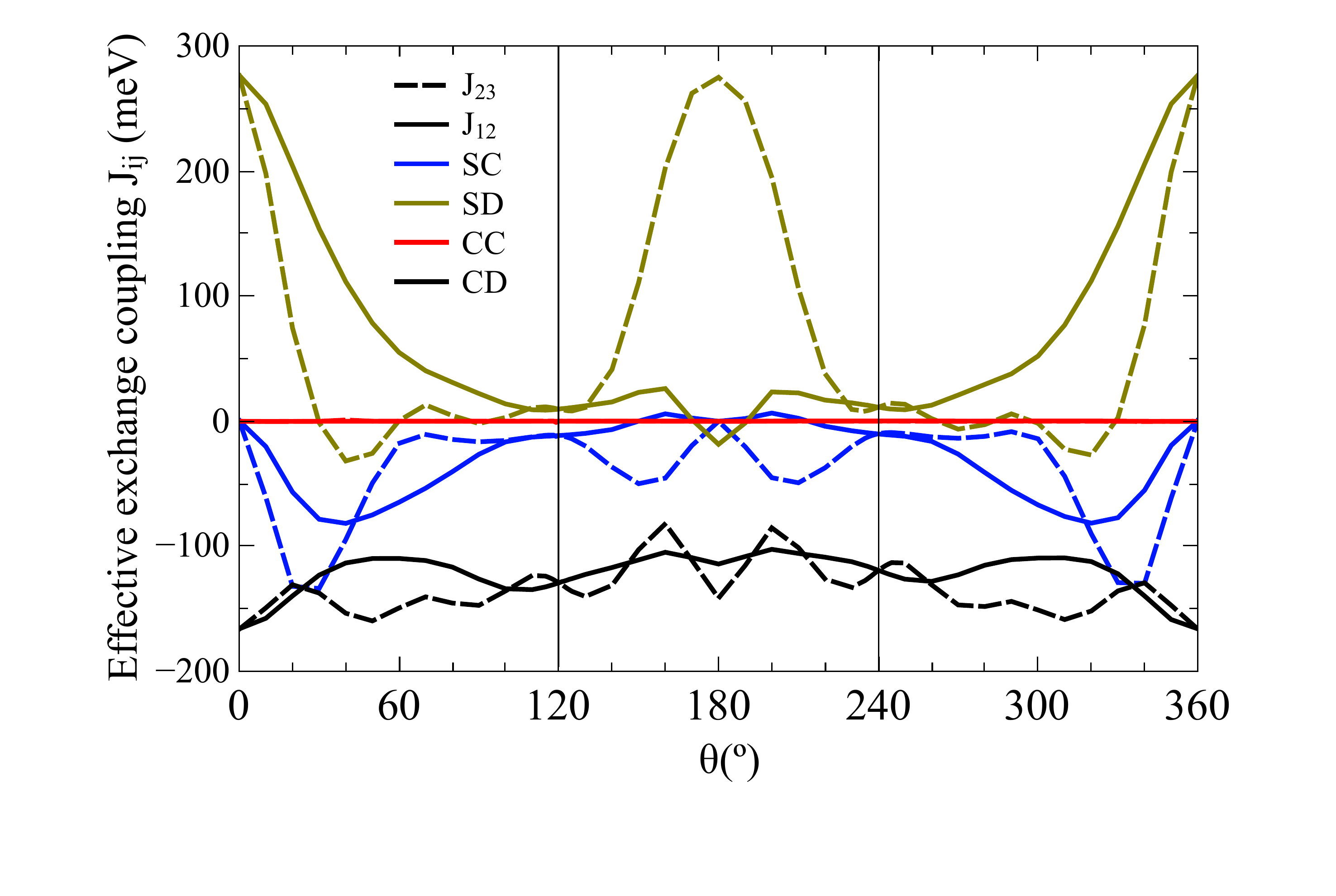}
\caption{Calculated values of the effective Heisenberg exchange of Mn$_3$Sn, between atoms of type 1 and 2 as well as atoms of type 2 and 3, as described in the main text. The results are shown as function of $\theta$, decomposed in the four terms of Eq.~(11) of the main text.}
\label{J12-resultat}
\end{center}
\end{figure} 

\subsection{Anisotropic interaction from the symmetric part of the exchange tensor}

In addition to the DM and Heisenberg types of interactions we also have the anisotropic but symmetric interaction, which is of tensorial form
\begin{align}
	\delta E^\mathrm{A}&=-2\sum_{ij}\delta\vec{m}_i\cdot\,A_{ij}\,\cdot\hat{m}_j\nonumber\\
	A_{ij}^{\alpha\beta}&=\frac{2}{\pi}\Im\int \mathrm{tr}\left(\Delta_i\, \vec{G}^{1}_{ij}\cdot\hat{\alpha}\,\Delta_j\,\vec{G}^{1}_{ji}\cdot\hat{\beta}-\right.\nonumber\\
	&\left.\Delta_i\, \vec{G}^{0}_{ij}\cdot\hat{\alpha}\,\Delta_j\,\vec{G}^{0}_{ji}\cdot\hat{\beta}\right)\,\mathrm{d}\varepsilon\label{defA}\,.
\end{align}
We do not here discuss this contribution further.

\subsection{Total energy differences}

If all variations are linear in angles,  we can readily integrate out a total energy difference
\begin{align}
	&\left\{E(\alpha)-E(0)\right\}=\nonumber\\
	4 &\int_0^\alpha\left\{J_{12}(\theta)\sin\theta+J_{23}(\theta)\sin 2\theta\right\}\mathrm{d}\theta+\nonumber\\
	   4 &\int_0^\alpha \left\{D^z_{12}(\theta)\,\cos\theta-D^z_{23}(\theta)\cos 2\theta\right\}\,\mathrm{d}\theta+\nonumber\\
	    4 &\int_0^\alpha\left\{A_{12}^{yy}(\theta)\,\sin\theta+\frac{1}{2}\left[A_{23}^{xx}(\theta)+A_{23}^{yy}(\theta)\right]\,\sin 2\theta\right\}\,
	   \mathrm{d}\theta\,.
	   \label{tote}
\end{align}
Note that there are eight independent effective interactions dependent on the angle $\theta$, and that they are summed up over all equivalent magnetic pairs of the full crystal. In Fig.~3 of the main part of this communication, it is clear how the DM interaction depends on the angle and as it is mainly the spin current part of the interaction, it follows the dependence of the strength of the intersite spin current in this co-planar magnetic system.

\subsection{Anisotropy}

The fact that magnetic space group is connected to the two dimensional irreducible representation, $E_{1g}$, implies that there exists a two-fold degeneracy in the absence of spin-orbit coupling.  In this case a uniform $90^\circ$ spin rotation around the $z$ axis gives a magnetic state that is degenerate with the one in Eq.~(2) of the main text. Any linear superposition of these two magnetic states are degenerate, but the SOC lifts this degeneracy, leading to a tiny anisotropy, calculated here with ELK package ({\it elk.sourceforge.net}) to be of the order 3 $\mu$eV.	

\subsection{Non-collinear counterpart to the spin-orbit coupling related effects.}

As discussed in the main paper, we suggest that several effects normally thought of as being caused by spin-orbit coupling can be complimentary to, or even dominated by, non-relativistic effects, e.g. due to a non-collinear arrangement of the spin texture. As discussed in the main part of the paper, this can lead to significant charge- and spin-currents, that are induced by the non-collinear spin-texture. 
This implies that in the field of spintronics and ultra-fast demagnetisation, spin-relaxation effects explained by Elliott-Yafet \cite{elliot,yafet} or D’yakonov-Perel’ \cite{dyakonov1972spin} mechanisms can have a significant contribution from non-relativistic effects.  Other related phenomena typically associated with spin-orbit coupling, that can be dominated by non-collinearity, is the magnetic anisotropy, the Gilbert damping parameter, the spin hall effect (SHE) as well as Rashba-like effects.
%

 


\end{document}